\documentclass[twocolumn,showpacs,superscriptaddress,pra]{revtex4}
\usepackage{graphicx}
\usepackage{amsmath}
\usepackage{amsthm}
\usepackage{amssymb}
\usepackage{latexsym}
\usepackage{array}
\usepackage{float}
\usepackage{amsfonts}
\usepackage{mathrsfs}
\usepackage{verbatim}

\begin{document}

\title{Procedures for realizing an approximate universal NOT gate}

\author{Jeongho Bang}
\affiliation{Center for Macroscopic Quantum Control \& Department of Physics and Astronomy, Seoul National University, Seoul, 151-747, Korea}
\affiliation{Department of Physics, Hanyang University, Seoul 133-791, Korea}

\author{Seung-Woo Lee}
\affiliation{Center for Macroscopic Quantum Control \& Department of Physics and Astronomy, Seoul National University, Seoul, 151-747, Korea}

\author{Hyunseok Jeong}
\affiliation{Center for Macroscopic Quantum Control \& Department of Physics and Astronomy, Seoul National University, Seoul, 151-747, Korea}

\author{Jinhyoung Lee}
\affiliation{Center for Macroscopic Quantum Control \& Department of Physics and Astronomy, Seoul National University, Seoul, 151-747, Korea}
\affiliation{Department of Physics, Hanyang University, Seoul 133-791, Korea}

\received{\today}

\begin{abstract}
We consider procedures to realize an approximate universal NOT gate in terms of average fidelity and fidelity deviation. The average fidelity indicates the optimality of operation on average, while the fidelity deviation does the universality of operation. We show that one-qubit operations have a sharp trade-off relation between average fidelity and fidelity deviation, and two-qubit operations show a looser trade-off relation. The genuine universality holds for operations of more than two qubits, and those of even more qubits are beneficial to compensating imperfection of control. In addition, we take into account operational noises which contaminate quantum operation in realistic circumstances. We show that the operation recovers from the contamination by a feedback procedure of differential evolution. Our feedback scheme is also applicable to finding an optimal and universal operation of NOT.
\end{abstract}

\pacs{03.67.Lx, 03.67.Pp}

\maketitle

\newcommand{\bra}[1]{\left<#1\right|}
\newcommand{\ket}[1]{\left|#1\right>}
\newcommand{\abs}[1]{\left|#1\right|}
\newcommand{\expt}[1]{\left<#1\right>}
\newcommand{\braket}[2]{\left<{#1}|{#2}\right>}
\newcommand{\ketbra}[2]{\left|{#1}\left>\right<{#2}\right|}
\newcommand{\commt}[2]{\left[{#1},{#2}\right]}

\newcommand{\tr}[1]{\mbox{Tr}{#1}}

\section{introduction}

Quantum information offers advantages in variety tasks over classical counterparts, by virtue of fundamental properties of quantum physics \cite{Nielsen99}. Quantum theory imposes, on the other hand, certain restrictions on quantum tasks \cite{Pati02}. For example, an arbitrary quantum state cannot be cloned, called no-cloning theorem, so that the superluminal (i.e. faster than light) communication via entanglement is prohibited \cite{Wootters82,Dieks82}. Another quantum task of universal-NOT (U-NOT) that transforms an arbitrary input state to its orthogonal is also restricted by quantum theory, while its classical task NOT is perfectly realized by bit-flip operation \cite{Buzek99,Buzek00-1}. This is because U-NOT cannot be implemented by a unitary operation but by an anti-unitary operation, violating the conditions of trace preservation and complete positivity that a physical procedure obeys \cite{Hellwig70,Choi75,Kraus83}.

An approximate realization of U-NOT task can nevertheless be done by a physical operation assisted by ancillary qubits \cite{Buzek99,Buzek00-1}. The approximate operation is the most optimal when it yields the average fidelity $2/3$ \cite{Buzek99}. Such an optimal operation has extensively been studied for the last decade both theoretically \cite{Buzek99,Buzek00-1,Enk05} and experimentally \cite{Sciarrino07,Martini02,Sias03} in order to clarify capabilities and limitations of quantum information processing. In another perspective, the optimal operation of U-NOT is closely related to other important quantum tasks such as quantum cloning, quantum state estimation, and entanglement test \cite{Buzek99,Enk05,Sciarrino07,Gisin99,Lim11}. In particular, U-NOT is equivalent to the transposition by some unitary transformation \cite{Lim11,Gisin99}. This implies that the optimal operation of U-NOT enables to approximately test if a quantum state is entangled with negative partial transposition \cite{Peres96,Horodecki09}.

A physical operation has been evaluated in terms of the fidelity between its output state and the target of the task. In particular, the average fidelity over all possible input states has been employed as an optimality measure of the operation to the task. However the average fidelity itself tells nothing about universality of operation, the condition that the task is performed equally for all possible input states. In a theoretical side, universality can be imposed on a quantum operation by requiring it to result in an equal fidelity for all input states. On the other hand, such requirement is nontrivial in experiments where imperfections of control and noises by environment arise. It is thus desired to consider a measure to quantify the condition for the operation to be universal over all input states. As such a measure, we employ fidelity deviation which is defined by the standard deviation of fidelity over all possible input states.

In this paper, we consider procedures to realize an approximate universal NOT gate. For the purpose, we characterize its approximate operations in terms of average fidelity $F$ and fidelity deviation $\Delta$. In the characterization, it is shown that one-qubit operations have a sharp trade-off relation between $F$ and $\Delta$; two-qubit operations exhibit a less sharp trade-off relation, including the one-qubit relation as an upper bound. The genuine universality of $\Delta = 0$ holds for $n$-qubit operations with $(n-1)$ ancillary qubits if $n \ge 3$, whereas, no matter how many qubits are involved in, the optimality is bounded in the average fidelity of $2/3$. Nevertheless, the operations of more than $3$ qubits can be beneficial to get more universality against imperfection of control. We can easily find a quantum operation of U-NOT which has rather high fidelity deviation even though its average fidelity is very close to its maximum. Therefore, investigating the universality and the optimality is important in the realization of U-NOT. In addition, considering some realistic circumstances, we take into account operational noises which contaminate quantum operation once optimized. We find a case that such a polluted operation is far from the universality no matter how close its average fidelity is to the maximum of $2/3$. To protect an operation against operational noises, we suggest a feedback scheme of using a differential evolution, showing that our scheme recovers the operation from the contamination as far as the noises fluctuate slowly compared to the operation. It is discussed that our scheme of feedback is applicable to find an optimal operation of U-NOT with no {\em a priori} knowledge except the number of qubits.

This paper is organized as follows. In Sec.\,II, we introduce our approach for the optimality and the universality by employing average fidelity $F$ and fidelity deviation $\Delta$. Sec.\,III is devoted to analyses of quantum operations for task U-NOT on the two-dimensional space of ($F$, $\Delta$). In Sec.\,IV, we investigate effects of operational noises on optimal operation of U-NOT and we suggest a feedback scheme to cure the contaminated operation by noise. Remarks are given in Sec.\,V.

\section{Average fidelity and fidelity deviation}

A task is realized by a physical operation $\hat{O}$ that transforms an input state $\ket{\Psi}$ to its target state $\ket{\Psi_t}$. Some tasks can not be ideally realized as restricted by quantum laws, such as quantum cloning \cite{Wootters82}. It is thus desirable to find an approximate but optimal operation as close as possible to a given task.

The quality of a found operation $\hat{O}$ is commonly quantified by a quantum fidelity $f$, that is defined by the transition probability between the output state $\hat{O}\ket{\Psi}$ and the target state $\ket{\Psi_t}$ of an input state $\ket{\Psi}$:
\begin{eqnarray}
f[\Psi] = |\bra{\Psi_t}\hat{O}\ket{\Psi}|^2.
\end{eqnarray}
For a given operation, the fidelity varies in general on input states. If not, the quantum operation is said to be universal. The universality can thus be thought of as associated with the fluctuation of the fidelity $f$ over the input states. In that sense, by the universal operation, the task can be performed equally for all possible input states. However, in some realistic circumstances, it is difficult to achieve the universality even for the universal operation, due to noise(s) during the physical process and/or an unavoidable interaction with environment. It is thus necessary to introduce a measure to quantify how much the fidelity $f$ fluctuate depending on the input states and to examine whether to reduce it by altering experimental parameters. In this section, we propose to employ the average fidelity for quantifying the optimality of an operation and the fidelity deviation for quantifying the universality.

The average fidelity $F$ is defined as
\begin{eqnarray}
\label{eq:avgF}
F=\int{d\Psi}f[\Psi],
\end{eqnarray}
where the integral is over all possible input state $\ket{\Psi}$ and $d\Psi$ is a normalized Haar measure, $\int d\Psi=1$. The measure $F$ quantifies on average how well operation $\hat{O}$ transforms input state $\ket{\Psi}$ to their target states $\ket{\Psi_t}$; the value $F=1$ implies the task is perfectly performed for all possible inputs, while $F=1/2$ does a random task. The fidelity deviation ${\Delta}$ is given in terms of the standard deviation of $f$,
\begin{eqnarray}
\label{eq:stdD} \Delta = \left[\int{d\Psi}{f[\Psi]^2} - F^2\right]^{1/2}.
\end{eqnarray}
The fidelity deviation $\Delta$ has the minimum of $0$ if $f=F$ for all input states and otherwise it increases. Note
\begin{eqnarray}
\label{eq:limit_ug}
\Delta^2 \leq\int{d\Psi}{f}[\Psi] - F^2=F(1-F)\leq\frac{1}{4},
\end{eqnarray}
where the last equality holds when $F=1/2$. Thus, $\Delta$ is bounded as $0 \le \Delta \le 1/2$. By the two measures $F$ and $\Delta$, we characterize a task operation, as a point on the two-dimensional space of $(F, \Delta)$.

\section{Characterization of U-NOT operations on the space of $(F, \Delta)$ }

In this section, we consider approximate operations for implementing task U-NOT and place them on the space $(F, \Delta)$. An optimal operation of U-NOT was found among three-qubit operations \cite{Buzek99, Buzek00-1}. It is questioned whether there exist any operations of U-NOT among one- or two-qubit operations. We try to answer this question and generalize to arbitrary number of qubits.

\subsection{One-qubit operations for U-NOT} 

An input state of all possible pure states is given in the Bloch representation by
\begin{eqnarray}
\label{eq:input_st}
\hat{\rho}_{\mathrm{in}} = \ketbra{\Psi}{\Psi} = \frac{1}{2}(\hat{\openone}+\mathbf{a}^T \boldsymbol\sigma),
\end{eqnarray}
where $\hat{\openone}$ is the identity operator, $\mathbf{a}=(a_x, a_y, a_z)^T = \tr{(\boldsymbol\sigma^\dagger\hat{\rho}_\mathrm{in})}$ is a Bloch vector of unit norm in $3$-dimentional real vector space $\mathbb{R}^3$, and $\boldsymbol\sigma=(\hat{\sigma}_x, \hat{\sigma}_y, \hat{\sigma}_z)^T$ is a vector with its components being Pauli operators $\hat{\sigma}_j$ $(j=x,y,z)$. Note that all pure states are located on the surface of the Bloch sphere with $|\mathbf{a}|^2=1$. The task U-NOT is supposed to transform each input $\hat{\rho}_{\mathrm{in}}$ to its orthogonal state or the Bloch vector $\mathbf{a}$ to its antipodal $-\mathbf{a}$:
\begin{eqnarray}
\label{eq:target_st}
\hat{\rho}_{\mathrm{in}}^{\perp}=\ketbra{\Psi^{\perp}}{\Psi^{\perp}}=\frac{1}{2}(\hat{\openone}-\mathbf{a}^T\boldsymbol\sigma).
\end{eqnarray}
The state $\hat{\rho}_{\mathrm{in}}^{\perp}$ is the target state of task U-NOT. To find a physically realizable (approximate) operation for U-NOT, we consider an arbitrary one-qubit unitary operation, given by
\begin{eqnarray}
\label{eq:single_u}
\hat{U}=\exp{\left(-i\frac{\vartheta}{2}\,\mathbf{n}^T\boldsymbol\sigma\right)}=\cos{\frac{\vartheta}{2}}\,\hat{\openone} - i\sin{\frac{\vartheta}{2}}\,(\mathbf{n}^T\boldsymbol\sigma),
\end{eqnarray}
where $\mathbf{n}=(n_x, n_y, n_z)^T$ is a unit vector. The operation transforms $\hat{\rho}_{\mathrm{in}}$ to the output state,
\begin{eqnarray}
\label{eq:1q_process}
\hat{\rho}_{\mathrm{out}} = \hat{U}\hat{\rho}_{\mathrm{in}}\hat{U}^{\dagger} = \frac{1}{2}(\hat{\openone}+\mathbf{b}^T\boldsymbol\sigma),
\end{eqnarray}
where $\mathbf{b}=\mathbf{R}\,\mathbf{a}$ and $\mathbf{R}$ is a rotation matrix on $\mathbb{R}^3$. The operation $\hat{U}$ can be understood as a rotation $\mathbf{R}$, on the Bloch vector $\mathbf{a}$, of the angle $\vartheta$ along axis $\mathbf{n}$ \cite{Nielsen99,Tian04}. Note that the output state is also pure, {\em i.e.} $|\mathbf{b}|^2=1$.

The fidelity $f=\tr{\left(\hat{\rho}_\mathrm{in}^{\perp}\,\hat{\rho}_\mathrm{out}\right)}$ between the output state $\hat{\rho}_\mathrm{out}$ and the target state $\hat{\rho}_\mathrm{in}^{\perp}$ is given by \cite{Jozsa94},
\begin{eqnarray}
\label{eq:qf}
f[\mathbf{a}]=\frac{1}{2} \left(1 - \mathbf{a}^T\mathbf{R}\,\mathbf{a}\right).
\end{eqnarray}
The average fidelity over all possible input states or all Bloch vectors $\mathbf{a}$ on the Bloch surface is given by
\begin{eqnarray}
\label{eq:avgF_1q-1}
F_{1Q} = \int d\mathbf{a} \, f[\mathbf{a}] = \int d\mathbf{a} \, \frac{1}{2}\Big(1-\mathbf{a}^T\mathbf{R}\,\mathbf{a}\Big),
\end{eqnarray}
where $d\mathbf{a}$ is the (normalized) Haar measure over the surface of the Bloch sphere \cite{Bowdrey02}. The subscript ``$1Q$'' stands for one qubit.

Eq.\,(\ref{eq:avgF_1q-1}) is evaluated in a spherical coordinate system, where $\mathbf{a}^T = (\sin{\theta}\cos{\phi}, \sin{\theta}\sin{\phi}, \cos{\theta})$ and $d\mathbf{a}=\frac{1}{4\pi}\sin{\theta}d\theta d\phi$. The diagonal components $a_i R_{ii} a_i$ are integrated to be $\frac{1}{3}R_{ii}$, while the non-diagonal $a_i R_{ij} a_j$ are to vanish. Alternatively, one may utilize Schur's lemma (Sec.~$2.2$ in Ref \cite{Serre77}),
\begin{eqnarray}
\label{eq:schur1}
\left[\mathbf{O}_g\mathbf{X}\mathbf{O}_g^T\right]_G = \frac{1}{d}\tr{(\mathbf{X})}\, \mathbf{I}_d,
\end{eqnarray}
where $\mathbf{I}_d$ is an identity matrix in $d$-dimensional real vector space $\mathbb{R}^d$, $\mathbf{O}_g$ is an irreducible orthogonal representation of an element $g$ in a given group ${\it G}$, and $[F_g]_G$ denotes the average of $F_g$ over all elements $g \in G$: $[F_g]_G \equiv \int{dg \, F_g}$, where $dg$ is the (normalized) Haar measure such that $\int{dg}=1$. This holds for every matrix $\mathbf{X}$ on $\mathbb{R}^d$. By applying the lemma\,(\ref{eq:schur1}) to the group O(3) of $3$-dimensional rotations, the second term in Eq.\,(\ref{eq:avgF_1q-1}) results in
\begin{eqnarray}
\label{eq:int_1}
\int d\mathbf{a}\,\mathbf{a}^T\mathbf{R}\,\mathbf{a} = \frac{1}{3}\tr{(\mathbf{R})},
\end{eqnarray}
where we used the fact that every Bloch vector $\mathbf{a}$ is given by some rotation $\mathbf{R}$ from a certain reference $\mathbf{z}$, $\mathbf{a}=\mathbf{R}\mathbf{z}$, and the average over the Bloch sphere is equal to that over the rotation group O($3$), $\int d\mathbf{a} \, \mathbf{R}_{g(\mathbf{a})}^T \mathbf{R} \mathbf{R}_{g(\mathbf{a})} = \int dg \, \mathbf{R}_g^T \mathbf{R} \mathbf{R}_g = \frac{1}{3}\tr{(\mathbf{R})} \, \mathbf{I}_3$. Both methods result in
\begin{eqnarray}
\label{eq:avgF_1q}
F_{1Q} = \frac{1}{2}-\frac{1}{6}\tr{(\mathbf{R})},
\end{eqnarray}
where $\tr{(\mathbf{R})}=2\cos{\vartheta} + 1$ (see Appendix A). The maximum of $F_{1Q}$ is given to be $2/3$ when $\vartheta=\pi$ and the minimum is $0$ when $\vartheta=0$ or $2\pi$. It is remarkable that the maximal average fidelity of one-qubit operation is already equal to that of three-qubut operation for U-NOT \cite{Buzek99,Buzek00-1}. In the case, the found optimal operation is in the form of Eq.\,(\ref{eq:single_u}) with $\vartheta=\pi$ and $\mathbf{n}$ being an arbitrary unit vector.

\begin{figure}[t]
\includegraphics[width=0.4\textwidth]{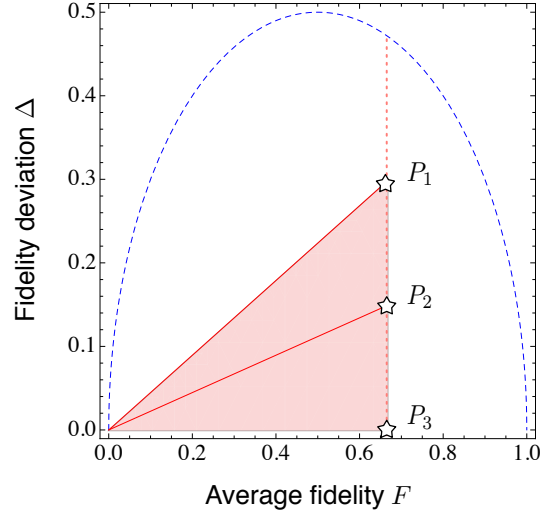}
\caption{(Color online) Accessible region of quantum operations for task U-NOT in terms of average fidelity $F$ and fidelity deviation $\Delta$. A one-qubit operation lies at a point on line $\overline{O P_1}$, where $O$ is the origin. An operation assisted by one quit (or two qubits) lies inside or on triangle $O P_1 P_2$ (or $OP_1P_3$). The (blue) dashed line stands for a mathematical boundary of quantum operations [see Eq.~(\ref{eq:limit_ug})].}
\label{fig:unot_sp}
\end{figure}

We investigate the fidelity deviation of one-qubit operations for task U-NOT. The square of the fidelity deviation is
\begin{eqnarray}
\label{eq:stdD_1q-1}
\Delta_{1Q}^2 &=& \int d\mathbf{a}\,f[\mathbf{a}]^2 - F_{1Q}^2, \nonumber \\
              &=& \frac{1}{4}\left[ \int d\mathbf{a}\,\left(\mathbf{a}^T\mathbf{R}\,\mathbf{a}\right)^2 - \frac{1}{9}\tr{(\mathbf{R})}^2 \right].
\end{eqnarray}
To evaluate $\Delta_{1Q}^2$ in Eq.\,(\ref{eq:stdD_1q-1}), we use a generalized identity of Schur's lemma in Eq.\,(\ref{eq:schur1}) to the tensor product of the two real vector spaces $\mathbb{R}^d \otimes \mathbb{R}^d$, given for each matrix $\mathbf{X}$ on $\mathbb{R}^d \otimes \mathbb{R}^d$,
\begin{eqnarray}
\label{eq:schur2}
\left[\left(\mathbf{O}_g\otimes\mathbf{O}_g\right)\mathbf{X}\left(\mathbf{O}_g^T\otimes\mathbf{O}_g^T\right)\right]_G=\alpha\mathbf{I}_{d^2} + \beta\mathbf{D} + \gamma\mathbf{P},
\end{eqnarray}
where
\begin{eqnarray}
\alpha &=& \frac{(d+1)\tr{(\mathbf{X})} - \tr{(\mathbf{X}\mathbf{D})} - \tr{(\mathbf{X}\mathbf{P})}}{d(d-1)(d+2)}, \nonumber \\
\beta &=& \frac{-\tr{(\mathbf{X})} + (d+1)\tr{(\mathbf{X}\mathbf{D})} - \tr{(\mathbf{X}\mathbf{P})}}{d(d-1)(d+2)}, \nonumber \\
\gamma &=& \frac{-\tr{(\mathbf{X})} - \tr{(\mathbf{X}\mathbf{D})} + (d+1)\tr{(\mathbf{X}\mathbf{P})}}{d(d-1)(d+2)}. \nonumber
\end{eqnarray}
Here, $\mathbf{P}$ is a swap matrix $\mathbf{P}\,(\mathbf{x}_i \otimes \mathbf{x}_j) = \mathbf{x}_j \otimes \mathbf{x}_i$, or equivalently,
$$
\mathbf{P}=\sum_{i,j=0}^{d-1}\left(\mathbf{x}_j\otimes\mathbf{x}_i \right) \left(\mathbf{x}_i\otimes\mathbf{x}_j \right)^T,
$$
and
$$\mathbf{D}=\left(\sum_{i=0}^{d-1} \mathbf{x}_i\otimes\mathbf{x}_i \right) \left( \sum_{j=0}^{d-1}\mathbf{x}_j\otimes\mathbf{x}_j \right)^T,
$$
where $\{\mathbf{x}_i\}$ is an orthonormal basis set in $\mathbb{R}^d$. Then, using the identity of Eq.\,(\ref{eq:schur2}), we rewrite the first term in Eq.\,(\ref{eq:stdD_1q-1}),
\begin{widetext}
\begin{eqnarray}
\label{eq:stdD_1q-int}
\int d\mathbf{a}\,\left(\mathbf{a}^T\mathbf{R}\,\mathbf{a}\right)^2
= \int d\mathbf{a}\,\left(\mathbf{a}\otimes\mathbf{a}\right)^T\left(\mathbf{R}\otimes\mathbf{R}\right)\left(\mathbf{a}\otimes\mathbf{a}\right)
=
\frac{1}{15} \Big[ \tr{(\mathbf{R}\otimes\mathbf{R})} + \tr{(\mathbf{R}\otimes\mathbf{R}\,\mathbf{D})}  + \tr{(\mathbf{R}\otimes\mathbf{R}\,\mathbf{P})} \Big],
\end{eqnarray}
\end{widetext}
where we used the similar reasoning below Eq.\,(\ref{eq:int_1}). Note that $\tr{(\mathbf{R}\otimes\mathbf{R})} = \tr{(\mathbf{R})}^2$, $\tr{(\mathbf{R}\otimes\mathbf{R}\,\mathbf{D})} = \tr{(\mathbf{R}\mathbf{R}^T)}=\tr{(\mathbf{I}_3)}=3$, and $\tr{(\mathbf{R}\otimes\mathbf{R}\,\mathbf{P})} = \tr{(\mathbf{R}^2)}$. Then Eq.\,(\ref{eq:stdD_1q-1}) is rewritten as
\begin{eqnarray}
\label{eq:delta_1q-m}
\Delta_{1Q}^2 &=& \frac{1}{4} \left\{ \frac{1}{15}\left[ \tr{(\mathbf{R})}^2 + 3 + \tr{(\mathbf{R}^2)} \right] - \frac{1}{9}\tr{(\mathbf{R})}^2 \right\} \nonumber \\
		&=& \frac{1}{5}\left[ \frac{1}{2} - \frac{1}{6}\tr{(\mathbf{R})} \right]^2 = \frac{1}{5}F_{1Q}^2,
\end{eqnarray}
where we used the relation $\tr{(\mathbf{R})}^2 - \tr{(\mathbf{R}^2)} = 2\,\tr{(\mathbf{R})}$ in Appendix A. The final form of the fidelity deviation $\Delta_{1Q}$ is given by
\begin{eqnarray}
\label{eq:1q_unot}
\Delta_{1Q} = \frac{1}{\sqrt{5}}F_{1Q}.
\end{eqnarray}
We note that this relation holds for arbitrary one-qubit operations as well as the optimal operations. This relation is represented by a segment $\overline{OP_1}$ in the space $(F, \Delta)$, as shown in Fig.~\ref{fig:unot_sp}. Eq.~(\ref{eq:1q_unot}) clearly shows the sharp trade-off relation between the conditions for one-qubit operations of task U-NOT to be optimal and universal: The larger the average fidelity, the larger the fidelity deviation. Thus, there is no one-qubit operation that satisfies both of universality and optimality as the condition $\Delta_{1Q}=0$ demands $F_{1Q}=0$, {\em i.e.} an identity operation, even though the maximal average fidelity is equal to that of three-qubit U-NOT.

\subsection{$n$-qubit operations assisted by ($n-1$) qubit(s)} 

\begin{figure}[b]
\includegraphics[width=0.37\textwidth]{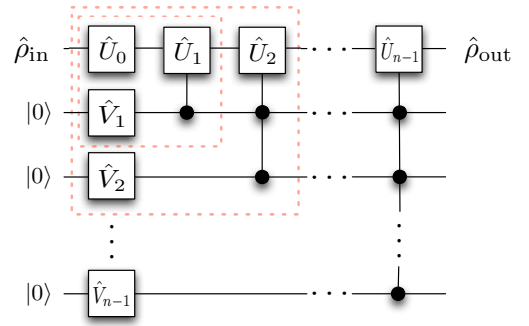}
\caption{(Color online) A quantum circuit for task U-NOT with $(n-1)$ ancillary qubits. The part inside the dashed small (large) box corresponds to one(two)-qubit assisting circuit.}
\label{fig:circuit_unot}
\end{figure}

We generalize one-qubit to $n$-qubit operations, by employing a specific type of logic circuits, as seen in Fig.\,\ref{fig:circuit_unot}. In the circuit, the first qubit is the system and the rest of ($n-1$) qubit(s) are ancillary. The case of $n=1$ was investigated in the previous sub-section. The circuit operation consists of local unitary $\hat{V}_j$ and conditional unitary $\hat{U}_j$. The local unitary operator $\hat{V}_j$ on the ancillary qubit $j$ is defined such that $\hat{V}_j\ket{0}_j = \sqrt{v_j}\ket{0}_j + \sqrt{1-v_j}\ket{1}_j$, where $v_j$ is a real number, satisfying $0 \le v_j \le 1$. The conditional unitary operator $\hat{U}_j$ acts on the system conditioned that ancillary qubits $k$ are in the state $\ket{1}_k$ for all $k \le j$.

When the circuit operation is applied on an input state $\ket{\Psi}$ of the system and the states $\ket{0}$ of the ancillary qubits, the output state $\hat{\rho}_\mathrm{out}$ becomes, partially tracing over the ancillary qubits,
\begin{eqnarray}
\label{eq:n-out}
\hat{\rho}_\mathrm{out} = \sum_{k=0}^{n-1} w_k \hat{W}_k \ket{\Psi}\bra{\Psi} \hat{W}_k^\dagger = \frac{1}{2}\left(\hat{\openone} + \mathbf{b}^T\boldsymbol\sigma \right).
\end{eqnarray}
Here, positive $w_k$ are functions of $v_j$'s,
\begin{eqnarray}
\label{eq:wk}
\left\{
\begin{array}{ll}
v_1 & (k=0), \\
(1-v_1)\cdots(1-v_k)v_{k+1} & (1 \le k \le n-2), \\
(1-v_1)\cdots(1-v_{n-2})(1-v_{n-1}) & (k=n-1),
\end{array}
\right. \nonumber \\
\end{eqnarray}
satisfying $\sum_{k=0}^{n-1}w_k=1$ and the unitary operators $\hat{W}_k$ are given by
\begin{eqnarray}
\label{eq:op_wk}
\hat{W}_k = \hat{U}_k\hat{U}_{k-1}\cdots\hat{U}_0.
\end{eqnarray}
Note that the output state $\hat{\rho}_\mathrm{out}$ is not necessarily a pure state, {\em i.e.} $|\mathbf{b}|^2 \le 1$, due to the entanglement created during the process. Nevertheless, it is remarkable that the circuit can be understood as a stochastic unitary map, in Eq.\,(\ref{eq:n-out}), characterized by the set of the local unitary operators $\hat{W}_k$ and the probability weights $w_k$. In other words, the circuit can be replaced by a stochastic circuit on the system that the unitary operation $\hat{W}_k$ is applied in the probability $w_k$. In the sense, the entanglement presented in the circuit is not necessarily demanded. The stochastic representation of operation in Eq.\,(\ref{eq:n-out}) reduces significantly the calculations in the characterization of the average fidelity and the fidelity deviation.

Consider the average fidelity $F_{nQ}$ (where the subscript ``$nQ$'' stands for $n$-qubit). The fidelity of the output state in Eq.\,(\ref{eq:n-out}) to the target state $\hat{\rho}_\mathrm{in}^\perp$ is given by
\begin{eqnarray}
\label{eq:qf_1qk}
f[\mathbf{a}]=\tr{(\hat{\rho}_\mathrm{in}^\perp \hat{\rho}_\mathrm{out})}=\sum_{k=0}^{n-1} w_k f_k[\mathbf{a}],
\end{eqnarray}
where $f_k[\mathbf{a}] = \left( 1-\mathbf{a}^T \mathbf{R}_k\mathbf{a} \right)/2$ and the rotation matrices $\mathbf{R}_k \in \mathbb{R}^3$ are associated with $\hat{W}_k$. Each $\hat{W}_k$ and thus $\mathbf{R}_k$ is given by the rotation angle $\vartheta_k$ and axis $\mathbf{n}_k$, as in Eq.\,(\ref{eq:single_u}). The average fidelity $F_{nQ}$ is given as
\begin{eqnarray}
\label{eq:avgFn}
F_{nQ} =\int d\mathbf{a}\left(\sum_{k=0}^{n-1} w_k f_k[\mathbf{a}] \right) = \sum_{k=0}^{n-1} w_k\,F_{1Q,k},
\end{eqnarray}
where $F_{1Q,k}$ is the average fidelity by a one-qubit operation $\hat{W}_k$, as in Eq.\,(\ref{eq:avgF_1q-1}). It is clear that $0 \le F_{nQ} \le 2/3$, as $F_{nQ}$ is a statistical mean of average fidelities of one-qubit operations. The maximum $F_\mathrm{nQ}=2/3$ is attained when $F_{1Q,k}$ are all equal to $2/3$, or equivalently $\vartheta_k = \pi$ for all $k$. This result holds for an arbitrary number of ancillary qubits. It thus seems that increasing the number of ancillary qubits does not improve the average fidelity or optimality for task U-NOT.

The square of the fidelity deviation $\Delta_{nQ}^2$ is given from Eqs.\,(\ref{eq:qf_1qk}) and (\ref{eq:avgFn}) as
\begin{eqnarray}
\label{eq:stdDn}
\Delta_{nQ}^2 &=& \int d\mathbf{a}\left(\sum_{k=0}^{n-1} w_k f_k[\mathbf{a}] \right)^2 - F_{nQ}^2 \nonumber \\
		&=& \sum_{k,l=0}^{n-1} w_k w_l C_{kl},
\end{eqnarray}
where $C_{kl}$ are elements of covariance marix $\mathbf{C}$, defined by
\begin{eqnarray}
\label{eq:cov}
C_{kl} = \int d\mathbf{a} f_k[\mathbf{a}] f_l[\mathbf{a}] - F_{1Q,k}F_{1Q,l}.
\end{eqnarray}
Note that $\mathbf{C}$ is symmetric, {\em i.e.} $C_{kl} = C_{lk}$. Each element of $\mathbf{C}$ is bounded, as shown in Appendix B, by
\begin{eqnarray}
\label{eq:cov_ineq}
\left\{
\begin{array}{ll}
C_{kk} = \Delta_{1Q,k}^2 \\
		\\
-\frac{1}{2}\Delta_{1Q,k}\Delta_{1Q,l} \le C_{kl} \le \Delta_{1Q,k}\Delta_{1Q,l},
\end{array}
\right.
\end{eqnarray}
where $\Delta_{1Q,k}$ is the fidelity deviation of one-qubit operation $\hat{W}_k$. The equality for the lower bound holds when the two rotation axes $\mathbf{n}_k$ and $\mathbf{n}_l$ are orthogonal to each other, {\em i.e.} $\mathbf{n}_k^T\mathbf{n}_l=0$, and the upper bound is reached when $\mathbf{n}_k$ and $\mathbf{n}_l$ are parallel or anti-parallel, {\em i.e.} $\mathbf{n}_k^T\mathbf{n}_l=\pm 1$. By Eq.\,(\ref{eq:cov_ineq}), the fidelity deviation $\Delta_{nQ}^2$ in Eq.\,(\ref{eq:stdDn}) is upper bounded,
\begin{eqnarray}
\label{eq:D_ub-nq}
\Delta_{nQ}^2 \le \left( \sum_{k=0}^{n-1} w_k \Delta_{1Q,k} \right)^2 = \frac{1}{5} F_{nQ}^2,
\end{eqnarray}
where we used Eqs.\,(\ref{eq:1q_unot}) and (\ref{eq:avgFn}). The equality holds when $\mathbf{n}_k^T\mathbf{n}_l=\pm 1$ for all pairs of $k \neq l$. The lower bound of $\Delta_{nQ}^2$ is given as
\begin{eqnarray}
\label{eq:D_lb-nq}
\Delta_{nQ}^2 &\ge& \sum_{k=0}^{n-1} w_k^2 \Delta_{1Q,k}^2 - \frac{1}{2}\sum_{k \neq l}^{n-1} w_k w_l \Delta_{1Q,k}\Delta_{1Q,l} \nonumber \\
		&\ge& \frac{3-n}{2}\sum_{k=0}^{n-1} w_k^2 \Delta_{1Q,k}^2,
\end{eqnarray}
where we used Eq.\,(\ref{eq:cov_ineq}) and the inequality, $\sum_{k \neq l}\left( w_k \Delta_{1Q,k} - w_l \Delta_{1Q,l}\right)^2 \ge 0$. The two equalities successively hold when $\mathbf{n}_k^T\mathbf{n}_l=0$ and $w_k \Delta_{1Q,k}=const$ for all pairs of $k \neq l$.

{\em Assisted by single ancillary qubit.} -- Based on the above results, let us consider two-qubit operations. The circuit is depicted inside the small dashed box in Fig.\,\ref{fig:circuit_unot}. The stochastic probabilities $w_k$ ($k=0,1$) are given by Eq.\,(\ref{eq:wk}), $w_0 = v_1$ and $w_1 = 1-v_1$. The average fidelity $F_{2Q}$ ranges from $0$ to $2/3$. When $\mathbf{n}_0^T\mathbf{n}_1 = 0$ and $w_0 \Delta_{1Q,0} = w_1 \Delta_{1Q,1}$, the lower bound of $\Delta_{2Q}^2$ in Eq.\,(\ref{eq:D_lb-nq}), $(1/2)\sum_{k=0,1} w_k^2 \Delta_{1Q,k}^2$, is attained and it is equal to $(1/4)(\sum_{k=0,1} w_k \Delta_{1Q,k})^2 = F_{2Q}^2/20$. Thus, the following inequalities hold,
\begin{eqnarray}
\label{eq:2q_unot}
\frac{1}{2\sqrt{5}}F_{2Q} \le \Delta_{2Q} \le \frac{1}{\sqrt{5}}F_{2Q}.
\end{eqnarray}
This implies a trade-off relation of $F_{2Q}$ and $\Delta_{2Q}$ for two-qubit operations, as represented by the triangle $O P_1 P_2$ in Fig.\,\ref{fig:unot_sp}. The trade-off relation in Eq.\,(\ref{eq:2q_unot}) is looser than one-qubit operations in the sense that for given average fidelity $F$ we can always find a two-qubit operation whose fidelity deviation is smaller than that of one qubit. The most optimal and universal operation is given when the operations $\hat{W}_0$ and $\hat{W}_1$ satisfy $\vartheta_0 = \vartheta_1 = \pi$ for their angles and $\mathbf{n}_0^T\mathbf{n}_1 = 0$ for their axes, and the stochastic probabilities $w_0 = w_1 = 1/2$. The fidelity deviation is reduced to $\Delta_{2Q} = 1/3\sqrt{5} \approx 0.15$ for the optimal operations of $F_{2Q} = 2/3$. We note that the circuit operations we have considered include all possible two-qubit operations and the current results hold in general as far as two qubits are involved.

{\em Assisted by two ancillary qubits.} -- Consider three-qubit operations, as shown in the large dashed box in Fig.\,\ref{fig:circuit_unot}. The local one-qubit unitary $\hat{V}_2$ and controlled-controlled-$\hat{U}_2$ operators are additionally employed for the task, and the stochastic probabilities $w_k$ are given by $w_0 = v_1$, $w_1 = (1-v_1)v_2$, and $w_2 = (1-v_1)(1-v_2)$. The lower bound $\Delta_{3Q}=0$ is reached when the three vectors $\mathbf{n}_k$ are mutually orthogonal and $w_k \Delta_{1Q,k} = const$, $\forall k$. Thus, we arrive at the trade-off relation, for three-qubit operations,
\begin{eqnarray}
\label{eq:3q_unot}
0 \le \Delta_{3Q} \le \frac{1}{\sqrt{5}}F_{3Q}.
\end{eqnarray}
This relation is represented by the triangle $O P_1 P_3$ in Fig.\,\ref{fig:unot_sp}. The most optimal and genuinely universal operation of $F_{3Q}=2/3$ and $\Delta_{3Q}=0$ is attained when the stochastic unitary operations $\hat{W}_k$ are given by their rotation angles $\vartheta_k = \pi$ and their axes mutually orthogonal $\mathbf{n}_k^T\mathbf{n}_l=0$ with $w_k = 1/3$ for all pairs of $k \neq l$. In terms of a stochastic map, the most optimal operation of U-NOT leads
\begin{eqnarray}
\hat{\rho}_\mathrm{in} \mapsto \hat{\rho}_\mathrm{out} &=& \frac{1}{3}\left( \hat{\sigma}_x\hat{\rho}_\mathrm{in}\hat{\sigma}_x + \hat{\sigma}_y\hat{\rho}_\mathrm{in}\hat{\sigma}_y + \hat{\sigma}_z\hat{\rho}_\mathrm{in}\hat{\sigma}_z \right) \nonumber \\
        &=& \frac{2}{3}\hat{\rho}_\mathrm{in}^{\perp} + \frac{1}{3}\hat{\rho}_\mathrm{in}.
\end{eqnarray}
This map is equivalent to the one found in Refs. \cite{Buzek00-1,Barnett10}.

The result in Eq.\,(\ref{eq:3q_unot}) still holds for more than $2$ ancillary qubits. Our analyses show that it is important to employ both indicators of the average fidelity and the fidelity deviation to evaluate a quantum operation of task U-NOT, because there exist operations whose average fidelity $F$ are close to $\frac{2}{3}$ but fidelity deviation $\Delta$ may be arbitrarily large, as implied by the line $\overline{P_1 P_3}$ of Fig.\,\ref{fig:unot_sp}. It is understood that such situation could be a case in experiments, as in Ref.\,\cite{Martini02,Sias03}, which will be discussed further in Sec. IV.

Before closing this section, remind that the three-qubit operations we have considered are the specific, as in the circuit, Fig.\,\ref{fig:circuit_unot}. One might question if there exist any three-qubit operation whose average fidelity is larger than $2/3$ when sacrificing the universality. This question is worth to investigate as the universality was assumed in the previous works \cite{Buzek99,Buzek00-1,Martini02,Sias03}. However, this is not the case. Consider an arbitrary three-qubit operation,
\begin{eqnarray}
\label{eq:uo_matrix}
\hat{U}_\mathrm{arb}=
\begin{pmatrix}
u_{00} && u_{01} && \cdots && u_{07} \\
u_{10} && u_{11} && \cdots && u_{17} \\
\vdots && \vdots && \ddots && \vdots \\
u_{70} && u_{71} && \cdots && u_{77} \\
\end{pmatrix}.
\end{eqnarray}
The average fidelity $F_{3Q}$ is a function of the matrix elements $u_{jk}$ ($j,k=0,1,\cdots,7$),
\begin{eqnarray}
\label{eq:avgF_ov_result}
F_{3Q} = \frac{2}{3} &-& \frac{1}{6}\Big(\abs{u_{00}+u_{44}}^2 + \abs{u_{10}+u_{54}}^2 \nonumber \\
        && + \abs{u_{20}+u_{64}}^2 + \abs{u_{30}+u_{74}}^2\Big).
\end{eqnarray}
The unitary condition, $\hat{U}_\mathrm{arb}\hat{U}_\mathrm{arb}^\dagger=\hat{\openone}$, leads to $0 \le F_{3Q} \le 2/3$. This proof can straightforwardly be generalized to arbitrary $n$ qubit operations.

There arises another question: Is there any advantage in using more than $2$ ancillary qubits? The answer is affirmative: Added qubits can be used to compensate or to absorb imperfection of operations if any. To see this, suppose that a stochastic operation $\hat{W}_k$ satisfies $\vartheta_k=\pi$ and $w_k=1/3$ for all $k=0,1,2$ and two rotation axes are not perfectly orthogonal with the angle $\frac{\pi}{2}-\alpha$ for small $\alpha$, say $\mathbf{n}_0^T\mathbf{n}_1=\mathbf{n}_0^T\mathbf{n}_2=0$ and $\mathbf{n}_1^T\mathbf{n}_2 = \cos{(\frac{\pi}{2}-\alpha)} \simeq \alpha$, neglecting higher order terms than $\alpha^2$. Then, even though the optimality is achieved with $F_{3Q}=2/3$, the universality is broken as $\Delta_{3Q} \simeq 2\alpha/3\sqrt{15} \neq 0$. In such circumstance, universality can be cured by extending the circuit from three to four qubits with $\hat{W}_3$ chosen such that $\vartheta_3=\pi$ and $\mathbf{n}_3$ is at the opposite direction to $\mathbf{n}_2$ by $\alpha$ on the plane $\mathbf{n}_1$-$\mathbf{n}_2$, that is, $\mathbf{n}_3^T\mathbf{n}_0=0$, $\mathbf{n}_3^T\mathbf{n}_1 = \mathbf{n}_2^T\mathbf{n}_1 \simeq \alpha$, and $\mathbf{n}_3^T\mathbf{n}_2 = \cos{2\alpha} \simeq 1-2\alpha^2$. By choosing the stochastic probabilities $w_0=w_1=1/3$, and $w_2=w_3=1/6$, then, the fidelity deviation becomes to $\Delta_{4Q} \simeq 0$ up to $\alpha^2$, while keeping $F_{4Q}\simeq F_{3Q}=2/3$. This example opens a possibility of recovering the universality without sacrificing any optimality when operations suffer from the imperfection.

\section{Feedback scheme to stabilize a quantum operation}

Implementing a quantum operation suffers from noise in realistic circumstance. To protect, we consider a feedback procedure with a differential evolution method, which is known as an efficient heuristic method for global optimization \cite{Storn97}. The adotion of such a feedback procedure is also beneficial when to find a quantum operation of itself. In this section, we introduce the differential evolution briefly and apply to the problem of finding an optimal operation of U-NOT among three-qubit operations. We show that the feedback scheme works so well that it consistently finds optimal operations of U-NOT, equivalent to the one in Ref.~\cite{Buzek99}. By introducing an operational noise which alters operational parameters unexpectedly, we show that the contaminated operation is cured by the feedback scheme as long as the noise fluctuates slowly.

\begin{figure}[t]
\includegraphics[angle=270,width=0.22\textwidth]{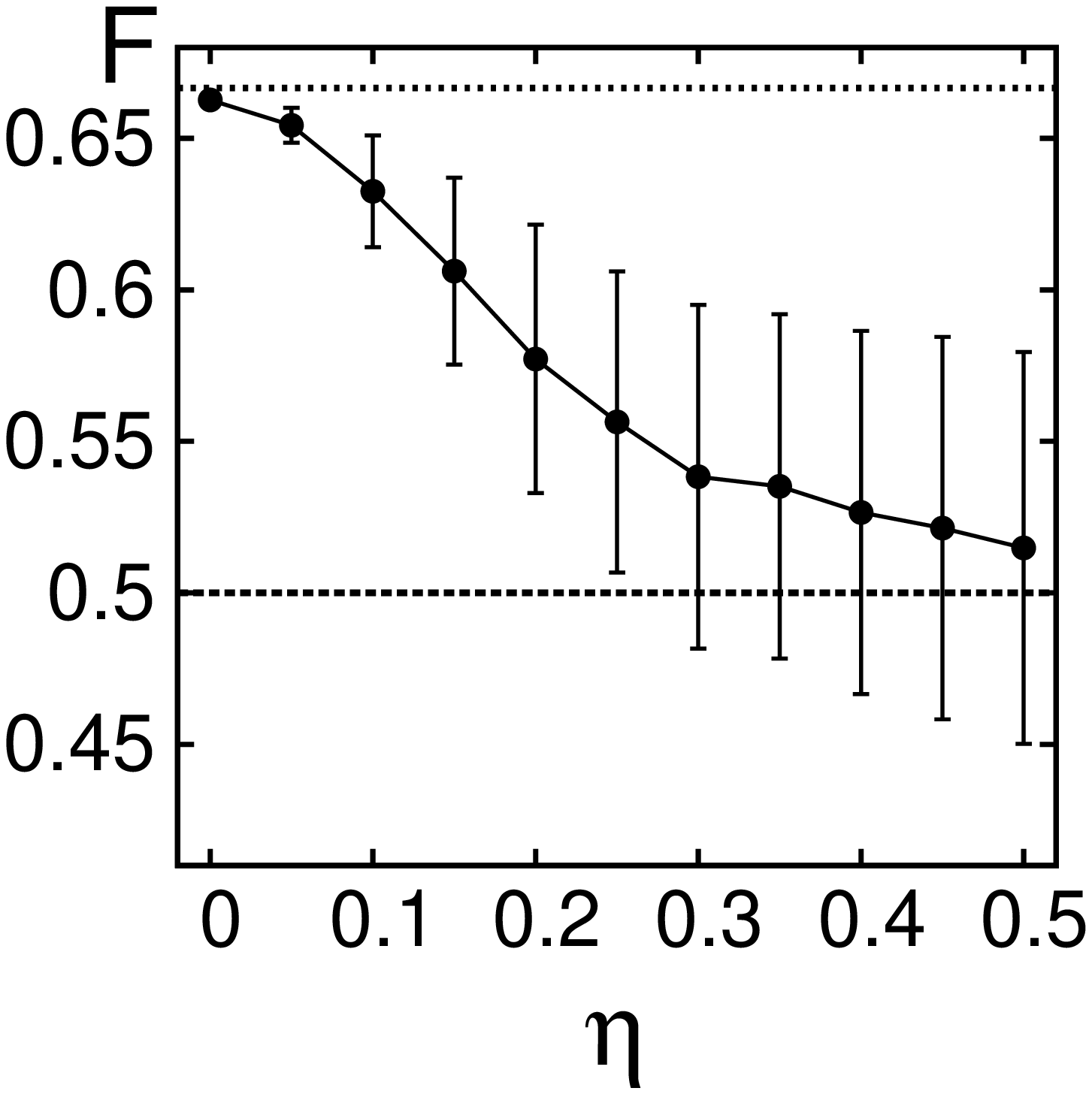}
\includegraphics[angle=270,width=0.22\textwidth]{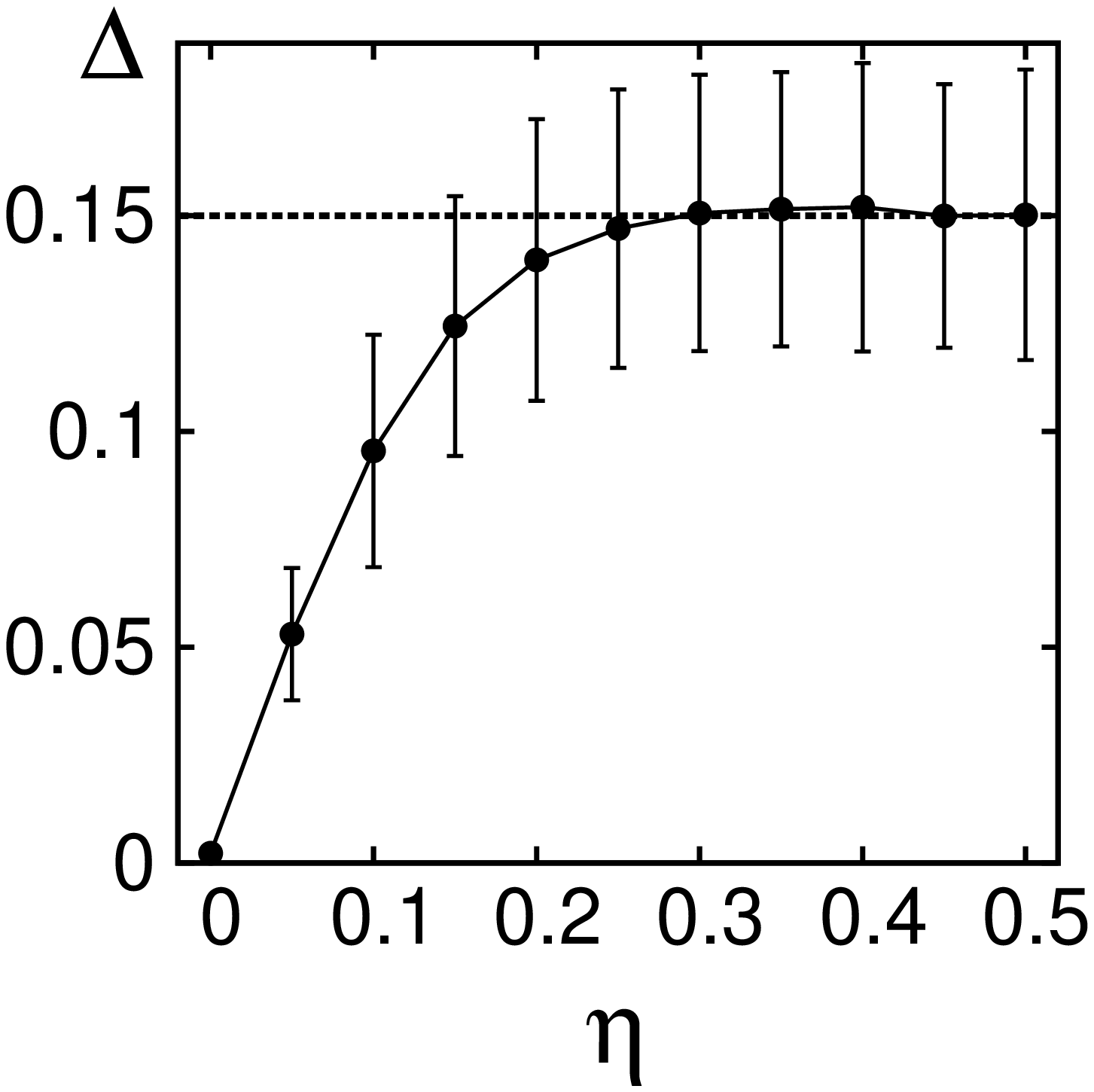}
\caption{Average fidelity $F$ (left) and fidelity deviation $\Delta$ (right) of the optimal U-NOT operation, when contaminated by operational noise of degree $\eta$. For each $\eta$, we perform 1000 simulations of Monte Carlo, averaging $F$ and $\Delta$ by the sample. Error bars are their standard deviations. Both of $F$ and $\Delta$ are degraded by the operational noise and such behaviors become conspicuous as increasing $\eta$. Dashed lines are of a random operation, $F_r = 1/2$ and $\Delta_r \simeq 0.150$, and dotted line in left graph is $F=2/3\simeq 0.666$.}
\label{fig:opt_data_err}
\end{figure}

\subsection{Effects of operational noise} 

Unitary operations on $d$-dimensional Hilbert space are parameterized by $(d^2-1)$-dimensional real vectors
$\mathbf{p}=(p_1,p_2,\cdots,p_{d^2-1})^T$ as
\begin{eqnarray}
\label{eq:unit_op}
\hat{U}(\mathbf{p}) = \exp{(-i\,\mathbf{p}^T\mathbf{G})},
\end{eqnarray}
where $\mathbf{G}=(\hat{g}_1, \hat{g}_2, \cdots,\hat{g}_{d^2-1})^T$ is a vector whose components are SU($d$) group generators $\hat{g}_j$ ($j=1,2,\cdots d^2-1$) \cite{Hioe81,Son04,Bang08}. Components of $\mathbf{p}$ are control parameters. Such a unitary operation can be realized by multiport beam splitters for an optical system \cite{Reck94} or pulse sequences for nuclear magnetic resonance system \cite{Lee00}. Based on the analysis of the previous section for U-NOT, we consider three-qubit unitary operations $\hat{U}(\mathbf{p})$ on $8$-dimensional Hilbert space with $8^2-1=63$ control parameters. Note that the number of control parameters can be reduced if any restriction on quantum operations are imposed, even though we assume no restrictions throughout this paper.

In the presence of noise, an operation $\hat{U}(\mathbf{p})$ turns to be imperfect with fluctuation of $\mathbf{p}$ \cite{Thomas11}. We choose a noise model in which fluctuation arises when dialing the control parameters $p_j$ such that
\begin{eqnarray}
\label{eq:err}
\mathbf{p} \rightarrow \mathbf{p} +\eta\,\boldsymbol\epsilon,
\end{eqnarray}
where $\boldsymbol\epsilon$ is a random stochastic error vector whose components $\epsilon_j$ are random between $-\pi$ and $\pi$. The factor $\eta$, normalized in $[0,1]$, stands for the degree of inaccuracy in control. This type of noise is supposed to occur in implementing $\hat{U}(\mathbf{p})$ and it is called an operational noise.

For U-NOT operations, we present the effects of operational noise on average fidelity $F$ and fidelity deviation $\Delta$ in Fig.\,\ref{fig:opt_data_err}. The average fidelity $F$ decreases and the fidelity deviation $\Delta$ increases as the degree of noise $\eta$ increases. That is, the performance of the operation is degraded, as expected. It is remarkable that for a small noise the average fidelity $F$ remains close to its maximum $2/3$ but the fidelity deviation $\Delta$ becomes rather large toward that of random operation, $\Delta_r = 1/{3\sqrt{5}}$. For instance, when $\eta = 0.1$, averaged over $1000$ samples, $F$ is $0.633 \pm 0.018$ which is about the value in an experiment \cite{Martini02,Sias03}, whereas $\Delta$ is rather high of $0.095 \pm 0.027$, compared to $\Delta_r \simeq 0.150$ (see Fig.~\ref{fig:opt_data_err}). In other words, $F$ is degraded by about $25\%$ from its maximum $3/2$ to that of random operation, $F_r = 1/2$, whereas $\Delta$ is increased by about $65\%$ to $\Delta_r$. The results support again the importance of the fidelity deviation in experimentally implementing a universal operation.

\subsection{Recovery from the contamination} 

Our differential evolution scheme of feedback is to find a set of values for control parameters $\mathbf{p}$ for an optimal and universal operation of NOT. The differential evolution algorithm follows \cite{Storn97}. To begin with, we account $N_\mathrm{pop}$ operations by which we are to develop approximate solutions. Then, we have $N_\mathrm{pop}$ parameter vectors $\mathbf{p}_i$ ($i=1,2,\cdots,N_\mathrm{pop}$), each of which consists of $63$ components $p_{j,i} \in [-\pi, \pi]$ ($j=1,2,\cdots,63$). All these $63 \times N_\mathrm{pop}$ parameters are chosen initially at random. [S.1] We generate $N_\mathrm{pop}$ mutant vectors ${\boldsymbol\nu}_i$ according to
\begin{eqnarray}
{\boldsymbol\nu}_i = \mathbf{p}_{a} + D \left(\mathbf{p}_{b} - \mathbf{p}_{b} \right),
\end{eqnarray}
where we randomly selected $a$, $b$, and $c$ among $N_\mathrm{pop}$ parameter vectors as far as they are mutually different. The free parameter $D$, called a differential weight, is a real and constant number we choose. [S.2] After that, the parameter vectors $\mathbf{p}_i=(p_{1,i}, p_{2,i}, \cdots, p_{63,i})^T$ are reformed to trial vectors $\boldsymbol\tau_i=(\tau_{1,i}, \tau_{2,i}, \cdots, \tau_{63,i})^T$ by the following rule: For each $j$,
\begin{eqnarray}
\label{eq:crossover}
\left\{
\begin{array}{ll}
\tau_{j,i} \leftarrow p_{j,i} & ~~\mathrm{if}~r_j > \mathrm{CR},\\
\tau_{j,i} \leftarrow \nu_{j,i} & ~~\mathrm{otherwise}, \\
\end{array}
\right.
\end{eqnarray}
where $r_j \in [0, 1]$ is a randomly generated number and the crossover rate $\mathrm{CR}$ is another free parameter we choose in $[0,1]$. [S.3] Lastly, the trial vector ${\boldsymbol\tau}_i$ is taken to be $\mathbf{p}_i$ for the next iteration if it yields a larger fitness value than ${\mathbf{p}}_i$, and otherwise ${\mathbf{p}}_i$ is retained. Here the fitness $\xi$ of a given operation $\hat{U}(\mathbf{p})$ is defined by
\begin{eqnarray}
\xi = F - \Delta,
\end{eqnarray}
where $F$ and $\Delta$ are the average fidelity and fidelity deviation for $\hat{U}(\mathbf{p})$, respectively.
It tells us how fit $\hat{U}(\mathbf{p})$ is to an optimal and universal operation of NOT \footnote{One might take another function of fitness. For instance, we tried $\xi = F (1-\Delta)$ but the overall tendency did not change much. This example does not exclude existence of a certain fitness function which would lead to very different behaviors.}. The steps [S.1]-[S.3] are repeated until the maximum iterations.

\begin{figure}[t]
\includegraphics[angle=270,width=0.23\textwidth]{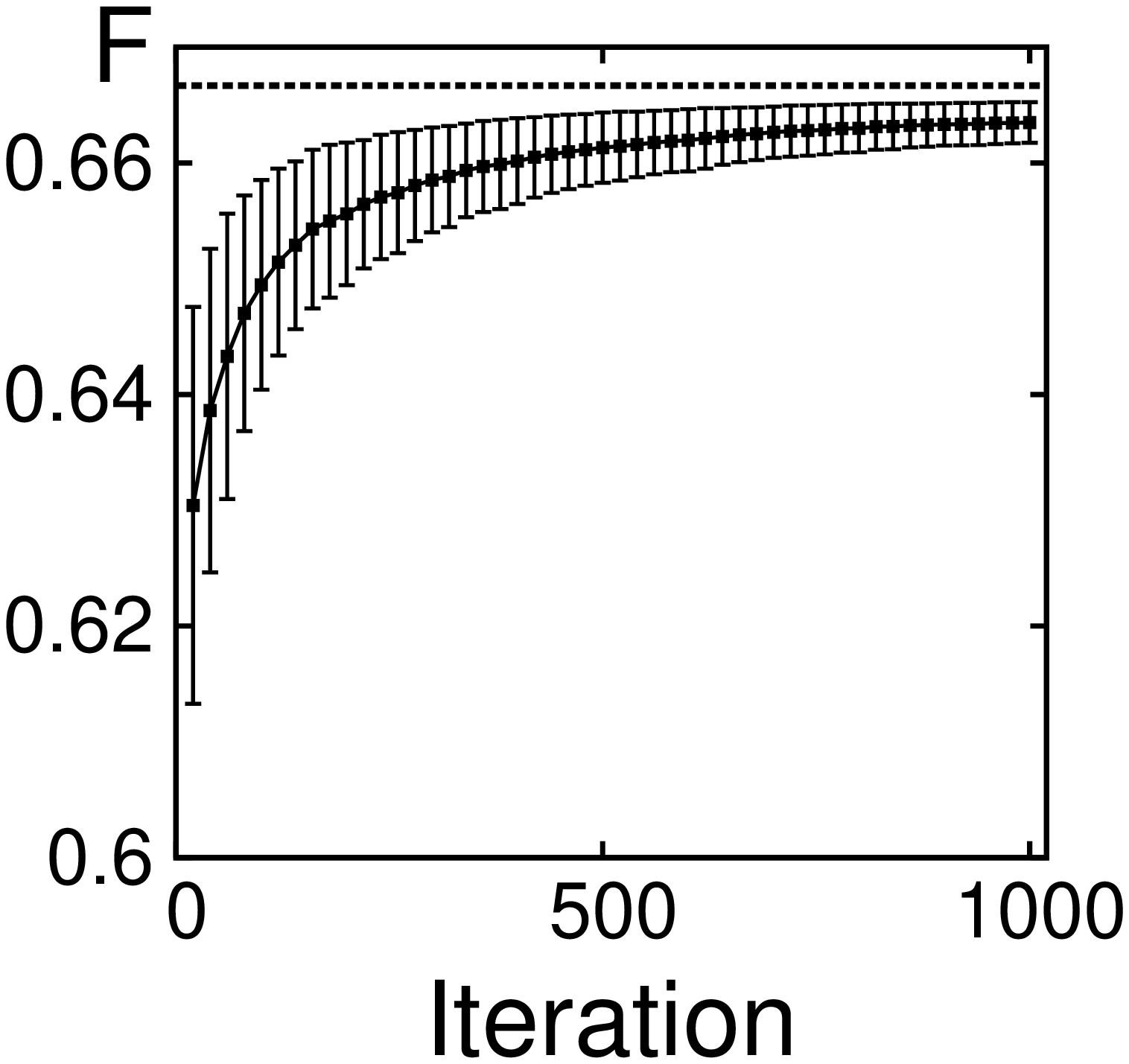}
\includegraphics[angle=270,width=0.23\textwidth]{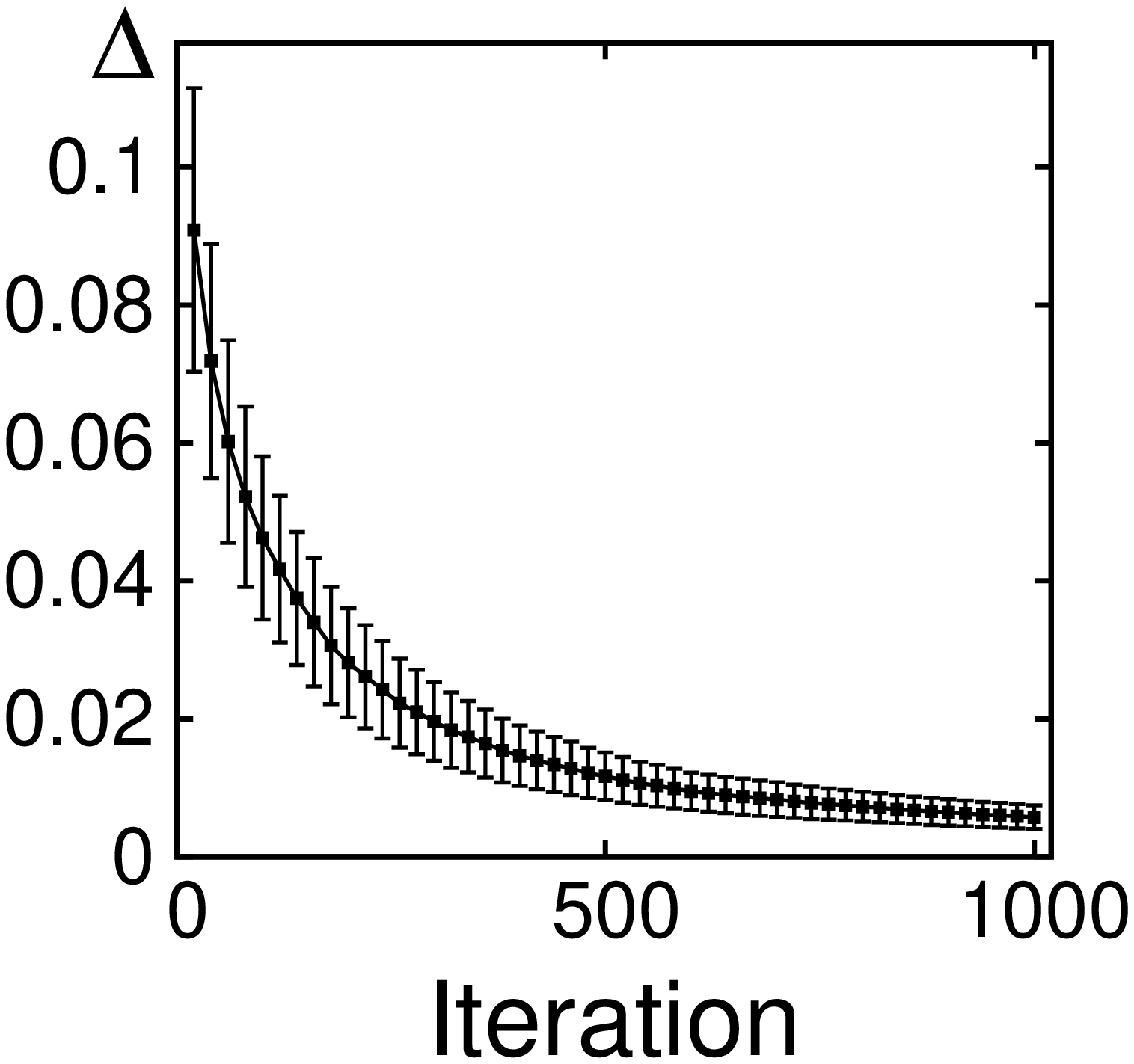}
\caption{Searching an optimal and universal operation of NOT by our feedback scheme of differential evolution in terms of average fidelity $F$ (left) and fidelity deviation $\Delta$ (right). For each iteration, we perform $1000$ Monte-Carlo simulations in averaging $F$ and $\Delta$ with their error bars. As iterating, $F$ and $\Delta$ steadily approach to their ideal optima, $F=2/3$ (dashed line) and $\Delta = 0$. We obtain $F = 0.663 \pm 0.002$ and $\Delta=0.006 \pm 0.002$ in $1000$ iterations.}
\label{fig:ga_data}
\end{figure}

We perform Monte-Carlo simulations. In the simulation, we take $N_\mathrm{pop}=10$, and the free parameters $D=0.1$ and $\mathrm{CR}=0.03$ which optimize our simulation. At every iteration, the fitnesses of all the operations are evaluated to select suitable parameters $\mathbf{p}_i$ for the next iteration, as described in [S.3]. We terminate the feedback procedure on $1000$ iterations. Fig.\,\ref{fig:ga_data} presents the average fidelity $F$ and the fidelity deviation $\Delta$ of the best among $N_\mathrm{pop}$ operations at every $20$ iterations. Both $F$ and $\Delta$ are statistically averaged by $1000$ trials of simulations. As seen in Fig.\,\ref{fig:ga_data}, $F$ converges to its ideal maximum $2/3\approx 0.667$ and $\Delta$ also converges to zero. We obtain $F = 0.663 \pm 0.002$ and $\Delta=0.006 \pm 0.002$ in $1000$ iterations. This result shows that our feedback scheme can be used to search the optimal and universal operation of NOT with no {\em a priori} knowledge on it, once the number of qubits is fixed \cite{Bang12}.

\begin{figure}[t]
\includegraphics[angle=270,width=0.23\textwidth]{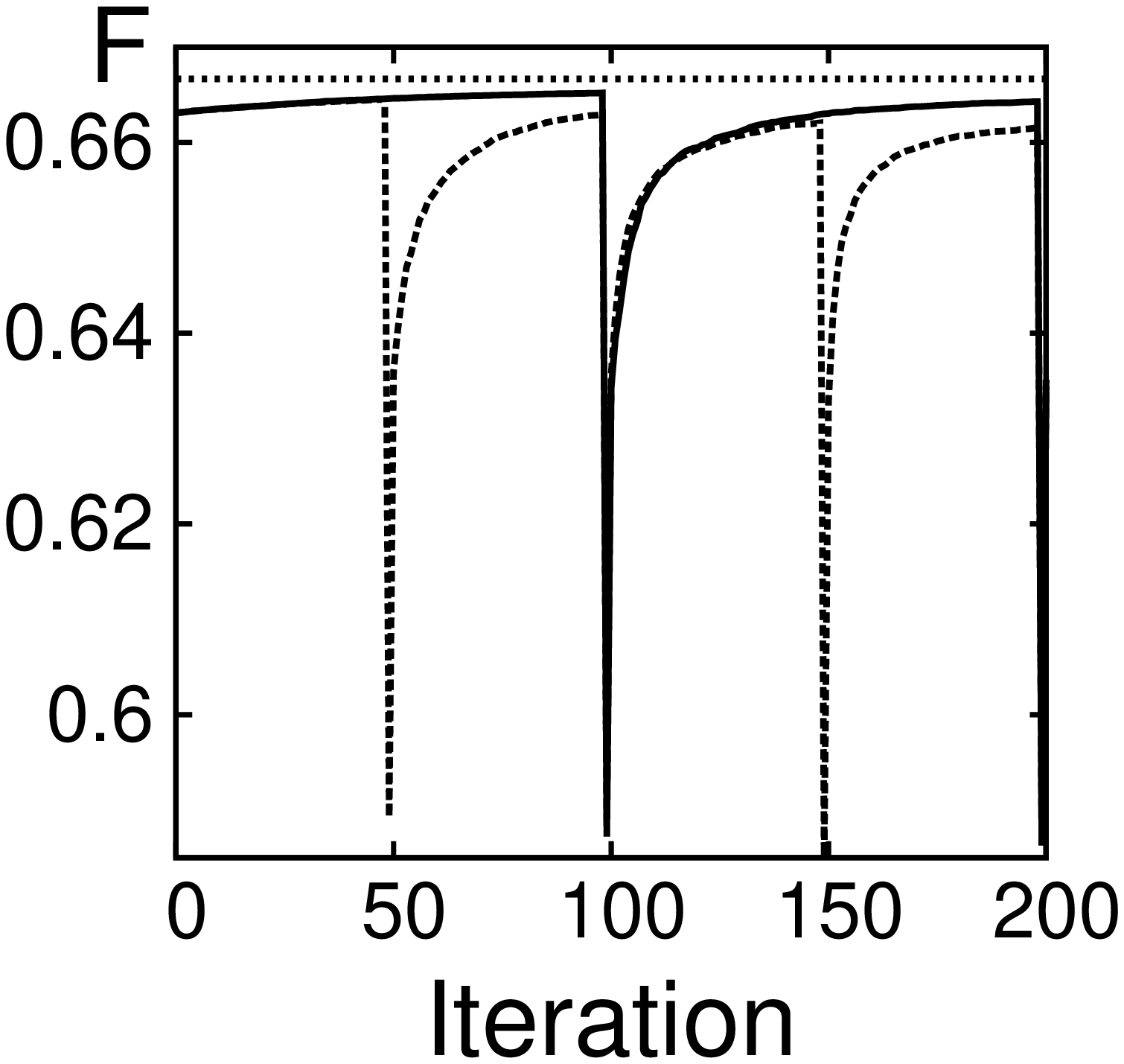}
\includegraphics[angle=270,width=0.23\textwidth]{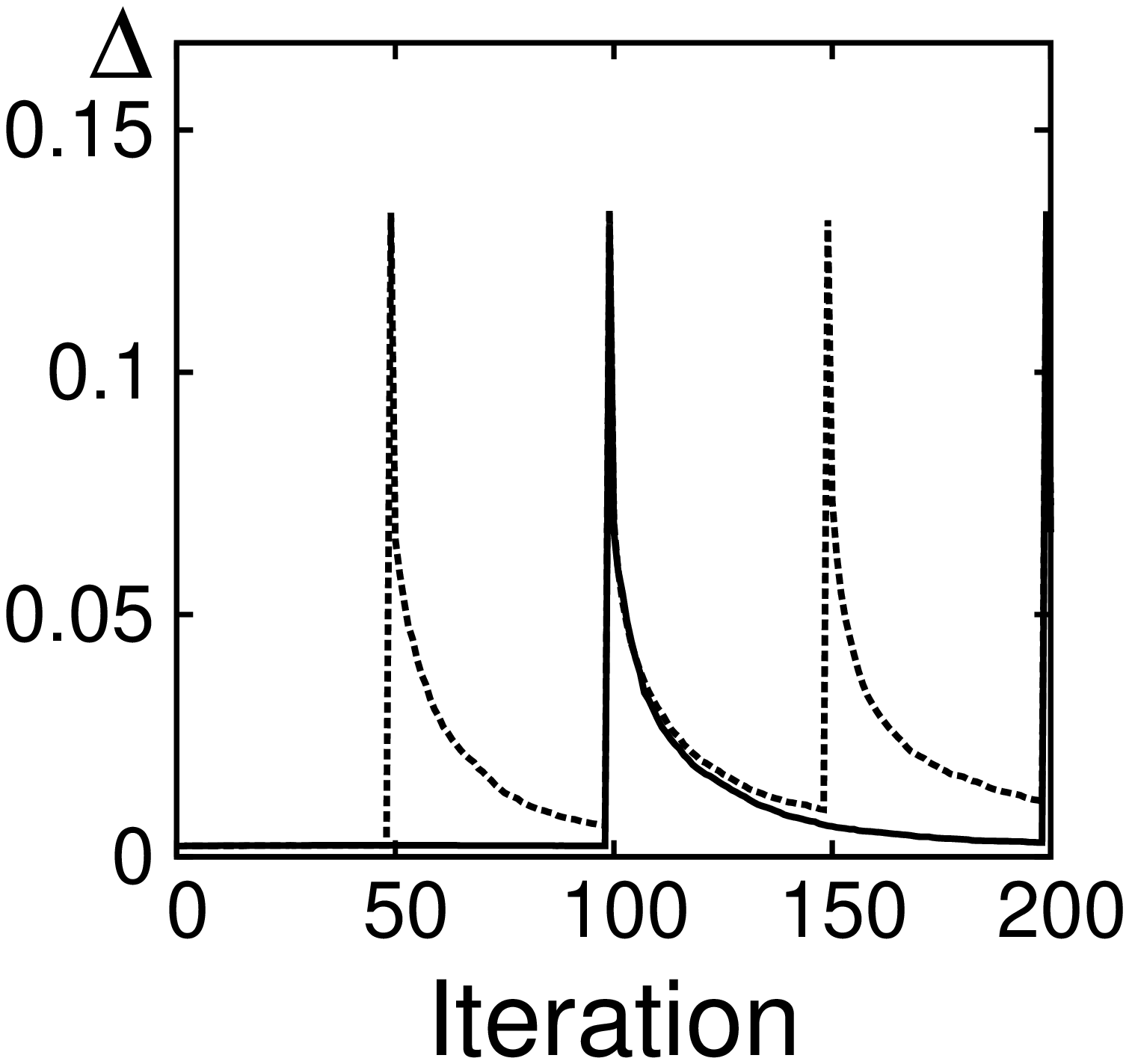}
\caption{Recovery of contaminated operation by the feedback scheme in terms of average fidelity $F$ (left) and fidelity deviation $\Delta$ (right). The operational noise is assumed to occur at every 50 (dashed line) or 100 (solid line) iterations with the noise degree $\eta =0.5$. The dotted line in left is $F=2/3\simeq 0.666$. Whenever the noise occurs, $F$ and $\Delta$ suddenly deteriorate close to those of random operation. As iterated, however, they steadily recover to those of the optimal and universal operation of NOT. The degree of recovery depends on the frequency of noise.}
\label{fig:comp_data_err}
\end{figure}

We also perform a Monte-Carlo simulation to test if our feedback scheme is able to recover the operation once contaminated by the operational noise. We assume that the noise fluctuates slowly compared to the operation, which is the case in most experiments for quantum tasks \cite{Viola99,Khaneja01}. Two cases are investigated that the abrupt fluctuation of noise occurs at every $50$ or $100$ iterations. Accounting the large degree of noise in Eq.~(\ref{eq:err}), we take $\eta=0.5$ for Fig.\,\ref{fig:comp_data_err}. Here, the operation initially optimized is polluted by the noise at every 50 or 100 iterations, on which the average fidelity $F$ and the fidelity deviation $\Delta$ suddenly deteriorate close to those of random operation. {\em As the feedback goes on, however, they steadily recover to those of the optimal and universal operation.} Note that the degree of recovery depends on both frequency and degree of noises, as one may expect.

\section{Remarks}

We have investigated procedures for realizing an approximate U-NOT gate by characterizing its approximate operations in terms of average fidelity $F$ and fidelity deviation $\Delta$. The average fidelity $F$ represented the optimality of operation on average, while the fidelity deviation $\Delta$ roughly does the fluctuation of the fidelity over the input states (reciprocally, ``universality''). The approximate operations could be characterized as a point on two-dimensional space of $(F,\Delta)$, by which way we analyzed the operation with respect to the optimality and the universality.

We showed that some of one-qubit operations can reach the average fidelity of $2/3$, the maximum limit attained by three-qubit optimal U-NOT, but lose their universality with high fidelity deviation. It was proved that there exists a quantum operation for arbitrary number of qubits such that it leads to the average fidelity of (but not larger than) $2/3$. The one-qubit operations showed a sharp trade-off relation, {\em i.e.}, a linear relation between $F$ and $\Delta$. Similar behaviors were observed for two-qubit operations, exhibiting a less sharp trade-off relation, {\em i.e.}, a triangular region on the space of ($F$, $\Delta$), which includes the one-qubit relation of trade-off as an upper bound. They could have the most universality of $\Delta=1/3\sqrt{5} \approx 0.15$. The genuine universality of $\Delta = 0$ was shown to hold for $n$-qubit operations with $(n-1)$ ancillary qubits as far as $n \ge 3$. Even though $3$-qubit operations suffice to optimally perform the U-NOT, it was shown that more-qubit operations can be beneficial against certain imperfections involved in. 

In some realistic circumstances, operational noises may arise in imperfect control of operation. The noises contaminate quantum operations even they are once optimized. We emphasized the existence of case that such a polluted operation is far from the universality no matter how its average fidelity is close to the maximum of $2/3$. This result supported again the importance of the fidelity deviation. In order to protect an operation against operational noises, we proposed a feedback scheme of using a differential evolution. It was shown that our scheme recovers the operation from the contamination, as far as the noises fluctuate slowly compared to the operation. We showed that our scheme of feedback is also applicable to find an optimal and universal operation of NOT with no {\em a priori} knowledge except the number of qubits.

We expect that our proposal of employing the measure of average fidelity and fidelity deviation will be applied to other universal quantum tasks such as cloning, teleportation, and inseparability test. Its modifications are eligible for partially universal tasks which involve a subset of states.

\section*{Acknowledgments}

We acknowledge the financial support of the National Research Foundation of Korea (NRF) grant funded by the Korea government (MEST) (No. 3348-20100018 and No. 2010-0015059).

\appendix

\section{$\tr{(\mathbf{R})}$ and $\tr{(\mathbf{R}^2)}$} 

We evaluate the traces of rotation matrices on three-dimensional real vector space $\mathbb{R}^3$. For the purpose, it is useful to represent a rotation matrix $\mathbf{R}$ in Rodrigues' form \cite{Alperin89}
\begin{eqnarray}
\label{eq:rot_rd}
\mathbf{R} = \mathbf{I}_3 - \sin\vartheta\,\mathbf{S} + \left(1-\cos\vartheta\right)\mathbf{S}^2,
\end{eqnarray}
where $\vartheta$ is the rotation angle and $\mathbf{I}_3$ is the identity matrix on $\mathbb{R}^3$. Here, $\mathbf{S}$ is the skew-symmetric matrix of cross product of the rotation axis $\mathbf{n}=(n_x, n_y, n_z)^T$, defined as
\begin{eqnarray}
\label{eq:skew_s}
{S}_{ij} = \sum_{k=\{x,y,z\}} \varepsilon_{ijk}\,n_k =
\begin{pmatrix}
0 && n_z && -n_y \\
-n_z && 0 && n_x \\
n_y && -n_x && 0 \\
\end{pmatrix},
\end{eqnarray}
where $\varepsilon_{ijk}$ is Levi-Civit\'a symbol. The squre of $\mathbf{S}$ in Eq.~(\ref{eq:skew_s}) is written as
\begin{eqnarray}
\label{eq:S2}
\mathbf{S}^2 = \mathbf{n}\otimes\mathbf{n}^T - \mathbf{I}_3.
\end{eqnarray}
From Eqs.\,(\ref{eq:skew_s}) and (\ref{eq:S2}) we obtain
\begin{eqnarray}
\label{eq:ssrs}
\tr{(\mathbf{I}_3)} = 3,~ \tr{(\mathbf{S})}=0, ~\mbox{and}~ \tr{(\mathbf{S}^2)}=-2.
\end{eqnarray}
Thus, the trace of $\mathbf{R}$ is given as
\begin{eqnarray}
\label{eq:trR_p}
\tr{(\mathbf{R})} = 2\cos{\vartheta} + 1,
\end{eqnarray}
which depends on the rotation angle $\vartheta$ but not the rotation axis $\mathbf{n}$. An alternative way to obtain $\mathrm{Tr} (\mathbf{R})$ is to find and sum eigenvalues of $\mathbf{R}$. As $\mathbf{R}$ has eigenvalues of $\{1, e^{\pm i\vartheta}\}$, their summation is equal to Eq.\,(\ref{eq:trR_p}).

We now prove the relation,
\begin{eqnarray}
\label{eq:tr_re}
\tr{(\mathbf{R})}^2 - \tr{(\mathbf{R}^2)} = 2\,\tr{(\mathbf{R})},
\end{eqnarray}
which was used in deriving Eq.\,(\ref{eq:delta_1q-m}). We first calculate $\mathbf{R}^2$ by using Eq.\,(\ref{eq:rot_rd}),
\begin{eqnarray}
\label{eq:R2}
\mathbf{R}^2 &=& \hat{\openone}_3 - 2\sin\vartheta\,\mathbf{S} + \left[ 2\left(1-\cos\vartheta\right) + \sin^2{\vartheta} \right]\mathbf{S}^2 \nonumber \\
		&& - 2\sin\vartheta \left(1-\cos\vartheta\right)\mathbf{S}^3 + \left(1-\cos\vartheta\right)^2 \mathbf{S}^4.
\end{eqnarray}
Noting that $\tr{(\mathbf{S}^3)}=0$ and $\tr{(\mathbf{S}^4)}=2$ and using Eqs.~(\ref{eq:ssrs}), we obtain the trace of $\mathbf{R}^2$,
\begin{eqnarray}
\label{eq:trR2}
\tr{(\mathbf{R}^2)} &=& 4\cos^2{\vartheta} - 1 \nonumber \\
		&=& \tr{(\mathbf{R})}^2 - 2\tr{(\mathbf{R})}.
\end{eqnarray}
This proves the relation in Eq.~(\ref{eq:tr_re}).

\section{Proof of Eq.\,(\ref{eq:cov_ineq})}  

In order to prove Eq.\,(\ref{eq:cov_ineq}), we recall the definition of $C_{kl}$ as in Eq.\,(\ref{eq:cov}). Substituting Eq.\,(\ref{eq:qf_1qk}) into Eq.\,(\ref{eq:cov}), we get
\begin{widetext}
\begin{eqnarray}
\label{eq:cov_d2}
C_{kl} &=& \frac{1}{4} \left[ \int d\mathbf{a} \left(\mathbf{a}^T \mathbf{R}_k \mathbf{a}\right) \left(\mathbf{a}^T \mathbf{R}_l \mathbf{a}\right) - \int d\mathbf{a} d\mathbf{b} \, \left(\mathbf{a}^T \mathbf{R}_k \mathbf{a} \right) \left(\mathbf{b}^T \mathbf{R}_l \mathbf{b} \right) \right] \nonumber \\
		&=& \frac{1}{4}\left[ \int d\mathbf{a}\,\left(\mathbf{a}\otimes\mathbf{a}\right)^T\left(\mathbf{R}_k\otimes\mathbf{R}_l\right)\left(\mathbf{a}\otimes\mathbf{a}\right) - \frac{1}{9}\tr{(\mathbf{R}_k)}\tr{(\mathbf{R}_l)} \right].
\end{eqnarray}
\end{widetext}
Let us define a couple of quantities,
\begin{eqnarray}
\label{eq:asrs}
A_1 &=& \tr{(\mathbf{R}_k\mathbf{R}_l^T + \mathbf{R}_k\mathbf{R}_l)}, \nonumber \\
A_2 &=& \tr{(\mathbf{R}_k)} \, \tr{(\mathbf{R}_l)}, \nonumber \\
A_3 &=& \tr{(\mathbf{R}_k)} + \tr{(\mathbf{R}_l)}.
\end{eqnarray}
Then we rewrite the first term in Eq.\,(\ref{eq:cov_d2}) by using Schur's lemma, as in Eq.~(\ref{eq:stdD_1q-int}), so that
\begin{eqnarray}
\label{eq:tmp_cov}
&& \int d\mathbf{a}\,\left(\mathbf{a}\otimes\mathbf{a}\right)^T\left(\mathbf{R}_k\otimes\mathbf{R}_l\right)\left(\mathbf{a}\otimes\mathbf{a}\right) \nonumber \\
&& = \frac{1}{15}\Big[ \tr{(\mathbf{R}_k)}\tr{(\mathbf{R}_l)} + \tr{(\mathbf{R}_k\mathbf{R}_l^T)} + \tr{(\mathbf{R}_k\mathbf{R}_l)} \Big] \nonumber \\
&& = \frac{1}{15} \left( A_1 + A_2 \right).
\end{eqnarray}
Then, $C_{kl}$ of Eq.\,(\ref{eq:cov_d2}) is reduced to
\begin{eqnarray}
\label{eq:tmp_cov2}
C_{kl} = \frac{3 A_1 - 2 A_2}{6^2 \times 5}.
\end{eqnarray}
Using Eq.\,(\ref{eq:rot_rd}), $A_1$ is explicitly calculated:
\begin{eqnarray}
\label{eq:cov_trin}
A_1 &=& 2\Big[ \cos{\vartheta_k}\cos{\vartheta_l} + \left(\cos{\vartheta_k} + \cos{\vartheta_l}\right) \nonumber \\
	&& + |\mathbf{n}_k^T\mathbf{n}_l|^2\left(1-\cos{\vartheta_k}\right)\left(1-\cos{\vartheta_l}\right) \Big].
\end{eqnarray}
Noting the last (third) term in Eq.~(\ref{eq:cov_trin}) is semi-positive, $A_1$ is upper bounded by
\begin{eqnarray}
\label{eq:A1_max}
A_{1,\mathrm{max}} = 4\cos{\vartheta_k}\cos{\vartheta_l} + 2 = A_2 - A_3 + 3.
\end{eqnarray}
It reaches the upper bound $A_{1,\mathrm{max}}$ when two rotation axes $\mathbf{n}_k$ and $\mathbf{n}_l$ are parallel or anti-parallel, i.e. $|\mathbf{n}_k^T\mathbf{n}_l|=1$. On the other hand, $A_1$ is lower bounded by
\begin{eqnarray}
\label{eq:A1_min}
A_{1,\mathrm{min}} &=& 2\left(\cos{\vartheta_k}\cos{\vartheta_l} + \cos{\vartheta_k} + \cos{\vartheta_l}\right) \nonumber \\
		&=& \frac{1}{2}\left( A_2 + A_3 - 3 \right).
\end{eqnarray}
The lower bound $A_{1,\mathrm{min}}$ is reached when $\mathbf{n}_k$ and $\mathbf{n}_l$ are ortogonal to each other or $\mathbf{n}_k^T\mathbf{n}_l=0$.

By substituting Eq.~(\ref{eq:A1_max}) into Eq.~(\ref{eq:tmp_cov2}), the upper bound of $C_{kl}$ is given by
$$
C_{kl,\mathrm{max}} = \frac{A_2 - 3A_3 + 9}{6^2\times 5} = \frac{[3-\tr{(\mathbf{R}_k)}][3-\tr{(\mathbf{R}_l)}]}{6^2\times 5}.
$$
Reminding of $\Delta_{1Q} = [3-\tr{(\mathbf{R})}]/6\sqrt{5}$ in Eq.~(\ref{eq:delta_1q-m}),
\begin{eqnarray}
\label{eq:cov_max}
C_{kl,\mathrm{max}} &=& \Delta_{1Q,k}\Delta_{1Q,l}.
\end{eqnarray}
Using Eq.~(\ref{eq:A1_min}), similarly, the lower bound of $C_{kl}$ is given by
\begin{eqnarray}
\label{eq:cov_min}
C_{kl,\mathrm{min}} &=& -\frac{1}{2}\Delta_{1Q,k}\Delta_{1Q,l}.
\end{eqnarray}
In case of $k=l$, $|\mathbf{n}_k^T\mathbf{n}_l|=1$ and thus
\begin{eqnarray}
C_{kl} = \Delta_{1Q,k}^2.
\end{eqnarray}

\end{document}